\documentclass[a4paper,fleqn,useAMS,usenatbib]{mnras}

\pdfoutput=1

\usepackage{mathptmx}

\usepackage[T1]{fontenc}
\usepackage{ae,aecompl}

\usepackage[dvips]{graphicx}
\usepackage{amsmath}
\usepackage{amsfonts}
\usepackage{amssymb}
\usepackage{comment}
\usepackage[dvipsnames]{xcolor}
\usepackage[%
        ]{hyperref}
        
\hypersetup{
	colorlinks=true,	
	urlcolor=MidnightBlue,		
	pdfpagelabels=true,
	hypertexnames=true,
	plainpages=false,
	naturalnames=true,
	pdftitle={Transit Origami},    
	pdfauthor={Kipping},     
	linkcolor=WildStrawberry,          
	citecolor=ForestGreen,        
}

\newcommand{\kepler}{{\it Kepler}}

\newcommand{\multi}{{\tt MultiNest}}
\newcommand{\forecaster}{{\tt forecaster}}
\newcommand{\luna}{{\tt LUNA}}

\newcommand{\wwwcoolworlds}{\href{https://github.com/CoolWorlds/OrigamiFigures}{https://github.com/CoolWorlds/TransitOrigamiBits}}

\title[Transit Origami]{
Transit Origami: A Method to Coherently Fold Exomoon Transits in Time Series Photometry 
}
\author[Kipping]{David Kipping$^{1}$\thanks{E-mail:
\href{mailto:dkipping@astro.columbia.edu}{dkipping@astro.columbia.edu}}\\
$^{1}$Dept. of Astronomy, Columbia University, 550 W 120th Street, New York NY 10027}

\date{Accepted 2021 July 7. Received 2021 July 4; in original form 2021 May 29}

\pubyear{2021}

\begin{document}
\label{firstpage}
\pagerange{\pageref{firstpage}--\pageref{lastpage}}
\maketitle

\begin{abstract}
One of the simplest ways to identify an exoplanetary transit is to phase fold a
photometric time series upon a trial period - leading to a coherent stack when
using the correct value. Such phase-folded transits have become a standard
data visualisation in modern transit discovery papers. There is no analogous
folding mechanism for exomoons, which would have to represent some kind of
double-fold; once for the planet and then another for the moon. Folding with
the planet term only, a moon imparts a small decrease in the surrounding
out-of-transit averaged intensity, but its incoherent nature makes it far less
convincing than the crisp stacks familiar to exoplanet hunters. Here, a new
approach is introduced that can be used to achieve the transit origami needed
to double fold an exomoon, in the case where a planet exhibits TTVs. This
double fold has just one unknown parameter, the satellite-to-planet mass ratio,
and thus a simple one-dimensional grid search can be used to rapidly identify
power associated with candidate exomoons. The technique is demonstrated on
simulated light curves, exploring the breakdown limits of close-in and/or
inclined satellites. As an example, the method is deployed on Kepler-973b, a
warm mini-Neptune exhibiting an 8 minute TTV, where  the possibility that the
TTVs are caused by a single exomoon is broadly excluded, with upper limits
probing down to a Ganymede-sized moon.
\end{abstract}

\begin{keywords}
planets and satellites: detection --- techniques: photometric
\end{keywords}

\section{Introduction}

The detection of a transiting planet within a photometric time series has
become a mature science after more than two decades of practice
\citep{charbonneau:2000,henry:2000} and many thousands of successes (see NASA
Exoplanet Archive; \citealt{akeson:2013}). Although in rare cases transits are
identified by eye (e.g. \citealt{wang:2013}), most detections have resulted
from automated algorithms - with the so-called ``Box Least Squares'' (BLS)
algorithm \citep{kovacs:2002} being perhaps the most well-known. Modifications
to this approach continue to be developed (e.g. TLS; \citealt{hippke:2019}),
but the core principle remains unchanged - fold a transit light curve upon a
trial period and examine the result for a coherent, short-duty cycle dip.

A folded transit light curve does not have higher signal-to-noise ratio (SNR)
than the original ensemble of $N$ transits. When compared to a single
(unfolded) transit, the folded transit representation decreases the noise
($\sigma$) at each phase point from $\sigma \to \sigma/\sqrt{N}$ (for Gaussian
statistics). Thus, the folded transit is $\sqrt{N}$ times more precise than a
single transit. In contrast, whilst the unfolded original time series has the
original noise $\sigma$, it contains $N$ times more transits than a single
transit. Thus, the ensemble, unfolded time series yields a transit precision
$\sqrt{N}$ times more precise than a single transit. In principle then, there
is no difference in overall SNR between the two\footnote{In practice, folding
can lead to slight reductions in SNR because the folded period can never be
perfect and thus the folded signal gets slightly distorted.}, but a
single high SNR folded transit is more visually impactful than a train of
lower SNR transits. Yet more, a folded transit immediately allows one to assess
whether the signal is coherent, as expected for a transiting planet (an
incoherent fold would higher scatter in the signal region caused by, for example,
varying transit depths). For these reasons, the folded transit light curve is a
common way of visualising new detections, to such an extent it has arguably
become a standardised accompanying visual.

Exomoon transits do not have an equivalent folding mechanism in the existing
literature. The basic challenge is that if one folds upon the planetary period,
the companion moon(s) will be in different relative phases in each epoch. Thus,
the moon transits are incoherent when phase-folded in the standard manner.
Despite being incoherent, their presence is still felt in the light curve. As
eluded to earlier, incoherent signals yield higher scatter and indeed this was
proposed by \citet{simon:2012} as a possible pathway towards their detection - 
seeking increased scatter in the regions surrounding the planetary transit.
In practice, \citet{hippke:2015} report that the act of masking transits during
the necessary light curve detrending means that the regions bracketing the
transits are least accurately detrended, which consequently imparts artificial
excess scatter into this region.

As an alternative, \citet{heller:2014} suggested focusing on measuring the mean
flux, rather than the scatter (which was dubbed the ``orbital sampling
effect'', OSE). Although the moon(s) indeed moves between
different positions yielding an incoherent phase-folded signal, it should
slightly decrease the average flux within the pre- and post-transit regions
surrounding the planetary transit imparting a broadly symmetric signature.
Crucially though, as discussed in \citet{teachey:2017}, a moon must spend less
than half of its time in each region (as it will spend a finite amount of time
overlapping with the planetary transit), which means that this averaging
process must be accomplished with an arithmetic mean rather than a median.
Medians offer a robust statistic than can ignore the influence of outlier
transits being stacked together, but because a median is the 50th percentile
position in a sorted list, it would remove the effect of an exomoon in the
outlined procedure. A remedy could be to mirror the phase-folded transits
prior to averaging, although this would sacrifice the distinctive symmetric
feature expected from this method. Despite the outlined progress then, exomoon
hunters still lack the kind of compelling visual of a planetary transit fold
that our exoplanet colleagues enjoy.

The true power of the conventional light curve fold is the fact that it
produces a coherent, clean stacked transit, with a well-resolved ingress,
egress and ``flat-bottom''. Although certainly astrophysical false-positives
exist here, the actual reality of some kind of photometric dip is far less
ambiguous than that present with the exomoon approaches discussed thus far. In
this work, a possible resolution to this is introduced, which subtly folds the
light curve in a novel way to recover a coherent moon transit. The paper is
organised as follows. In Section~\ref{sec:theory}, the underlying theory and
concept is introduced along with a suggested recipe. In Section~\ref{sec:SNR},
the expected SNR is estimated using the technique to aid the community in
quickly calculating detectability. In Section~\ref{sec:tests}, numerical tests
of the method are demonstrated, exploring the impact of inclination effects. In
Section~\ref{sec:application}, the method is applied to a real transiting
planet, Kepler-973b, where an exomoon is excluded down to Ganymede-radius.
Finally, the limitations of this method, as well as possible future development
and extensions, are discussed in Section~\ref{sec:discussion}.

\section{Introducing the Double-Fold}
\label{sec:theory}

\subsection{Predicting the position of an exomoon}
\label{sub:theory}

\begin{figure*}
\begin{center}
\includegraphics[width=18.0cm,angle=0,clip=true]{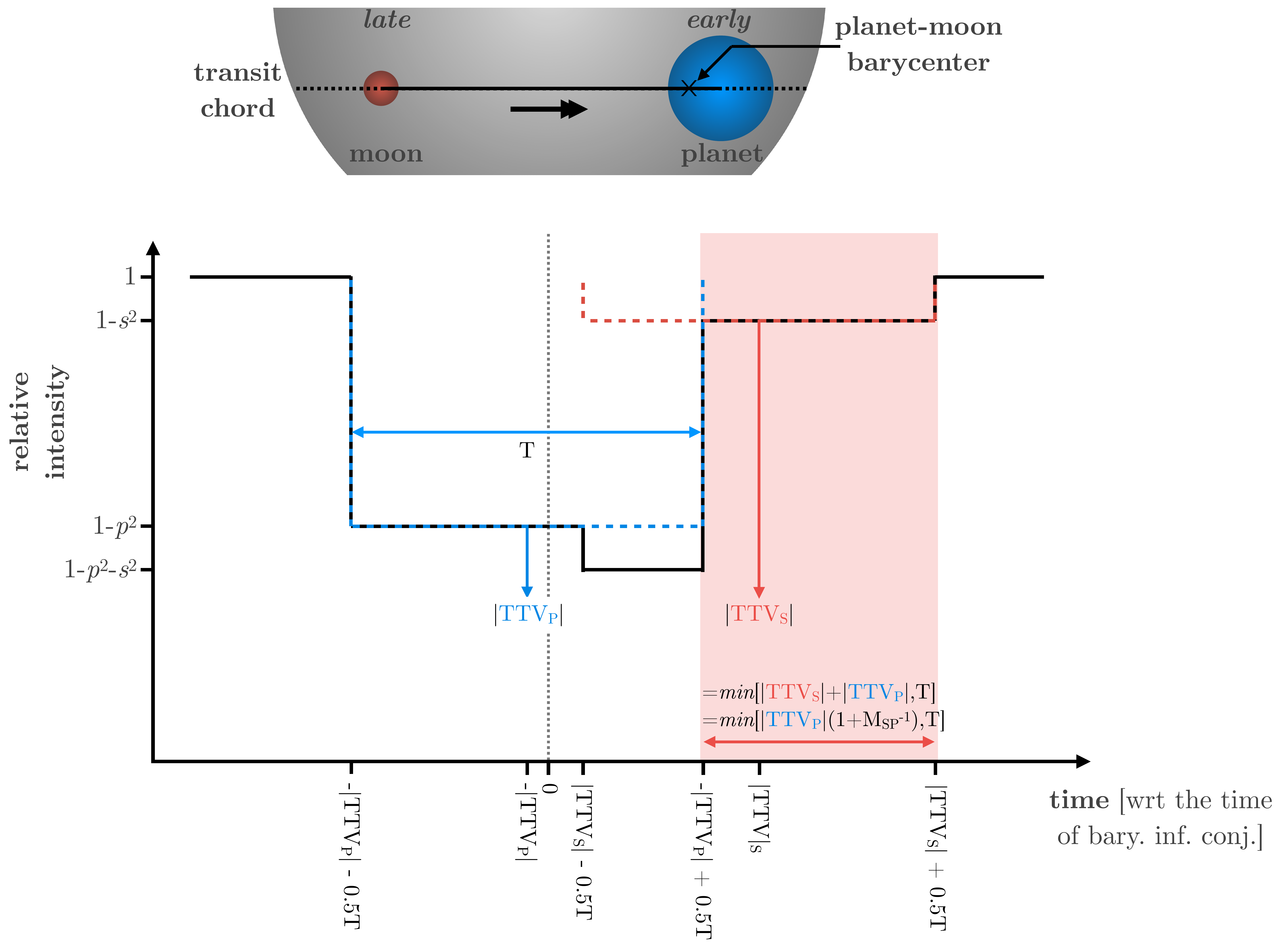}
\caption{
Simplified depiction of a planet-moon pair transiting a star. A coplanar
configuration is shown here for simplicity, but the concept generalises
to other configurations as discussed in the text. The fact that the planet
transits early in this instance (versus the barycentre), reveals that
the moon must be transiting late - modulo the unknown mass ratio,
$M_{SP}(\equiv M_S/M_P)$ term. Note that although the planet-moon pair appear
at quadrature, this is only a product of the projection shown and infact the
pair could be close to a conjunction. 
}
\label{fig:halfhouse}
\end{center}
\end{figure*}

To begin, consider the transit timing variations (TTVs) imparted upon a
planet by a single satellite, or indeed one large dominant moon.
A simple depiction of a planet-moon system transiting a star is shown in
Figure~\ref{fig:halfhouse} to guide the reader for what follows.

Following \citet{sartoretti:1999} and \citet{kipping:2009a}, this TTV effect
can be analytically derived by considering the offset of the planet away from
the planet-moon barycentre divided by the relative velocity of that barycentre
compared to the star. For a transiting planet with zero impact parameter, the
orbit lives entirely within the $\hat{X}$-$\hat{Y}$ plane, where $\hat{Z}$ is
the line-of-sight of the observer. Since transiting planets can necessarily
only have small inclinations (which is defined as an $\hat{X}$-rotation
conventionally), then most of motion during a transit occurs along the
$\hat{X}$-direction. This property greatly simplifies the derivation, and was
exploited explicitly in the derivation presented in \citet{thesis:2011}, such
that

\begin{align}
\mathrm{TTV}_P(t) &= \frac{(X_P - X_B)}{\big[\frac{\mathrm{d}X_B}{\mathrm{d}{t}}\big]_{\mathrm{inf.\,\,conj.}}},
\label{eqn:TTVformula}
\end{align}

where $\mathrm{TTV}_P(t)$ is the TTV effect experienced by the planet, $X_P$ is
the $\hat{X}$ Cartesian element of the planet's position (with respect to the
star), $X_B$ is the $\hat{X}$ Cartesian element of the planet-satellite
barycentre and ``inf. conf.'' refers to the instant of inferior conjunction.

In \citet{thesis:2011}, eccentric moon orbits are considered, but this study
simplifies the analysis by limiting the scope to circular orbit moons. The
close proximity of moons to their planets means that tidal forces rapidly
circularise satellites \citep{porter:2011,hamers:2018} and thus, unless they
are being continuously excited, one should generally expect circular orbits.
This also allows for considerable simplification in what follows.

The overall orbit around the star is often elliptical \citep{kane:2012,
xie:2016,vaneylen:2019}. Nevertheless, if one models a transit with a circular
orbit, the fits are essentially nearly identical to a full eccentric fit
\citep{barnes:2007}. The primary difference is that the inferred value of
$a/R_{\star}$ (or equivalently $\rho_{\star}$ if parametrised that way) is skewed
from the true value \citep{MAP:2012,dawson:2012}, which originates from the
differing orbital velocity. As a result, exoplanets are well-modelled and
explained by circular orbits, provided one understands that the resulting
$a/R_{\star}$ values should be treated as ``effective'' values.

Proceeding as described then, one may start by modifying the expressions of
\citet{thesis:2011} to the limiting case of circular orbits, to obtain

\begin{eqnarray}
	\mathbf{R}_B =
	\begin{bmatrix}
		X_B \\
		Y_B \\
		Z_B \\
	\end{bmatrix} =
	\begin{bmatrix}
		a_P \sin(n_P (t-t_{\mathrm{inf.\,\,conj.}})) \\
		a_P \cos(n_P (t-t_{\mathrm{inf.\,\,conj.}})) \cos i_P \\
		a_P \cos(n_P (t-t_{\mathrm{inf.\,\,conj.}})) \sin i_P
	\end{bmatrix}.
\label{eqn:Rb}
\end{eqnarray}

where $a_P$ is the semi-major axis of the planet-satellite barycentre around
the star, $n_P$ its mean motion and $i_P$ is the orbital inclination.

The satellite's orbit with respect to the planet-moon barycentre can be found
through a series of matrix rotations, and for circular orbits becomes

\begin{eqnarray}
	\mathbf{R}_{SB} =
	\begin{bmatrix}
		a_{SB} \cos\Omega_S \sin(n_S(t-\tau_S)) - \sin i_S \sin\Omega_S \cos(n_S (t-\tau_S)) \\
		a_{SB} \sin\Omega_S \sin(n_S(t-\tau_S)) + \sin i_S \cos\Omega_S \cos(n_S (t-\tau_S)) \\
		a_{SB} \cos i_S \cos(n_S (t-\tau_S))
	\end{bmatrix}.
\label{eqn:Rsb}
\end{eqnarray}

One may now write that $\mathbf{R}_S = \mathbf{R}_B + \mathbf{R}_{SB}$,
and also that $\mathbf{R}_P = \mathbf{R}_{B} - M_{SP} \mathbf{R}_{SB}$
where $M_{SP}(\equiv M_S/M_P)$ is the mass ratio between the satellite and the
planet. One can now evaluate Equation~(\ref{eqn:TTVformula}) to give

\begin{align}
\mathrm{TTV}_P(t) =& \frac{a_{SB} M_{SP}}{a_P n_P} \Big( \cos(n_S(t-\tau_S)) \sin i_S \sin\Omega_S \nonumber\\
\qquad& - \cos\Omega_S \sin(n_S(t-\tau_S)) \Big).
\label{eqn:TTVp}
\end{align}

It's not just the planet that experiences deviations in its position from the
barycentre. The satellite too experiences deviations - indeed much larger
deviations. The temporal offset of the satellite away from the time of inferior
conjunction, can be found using an analogous version of
Equation~(\ref{eqn:TTVformula}):

\begin{align}
\mathrm{TTV}_S(t) =& -\frac{a_{SB}}{a_P n_P} \Big( \cos(n_S(t-\tau_S)) \sin i_S \sin\Omega_S \nonumber\\
\qquad& - \cos\Omega_S \sin(n_S(t-\tau_S)) \Big).
\label{eqn:TTVs}
\end{align}

In fact, it is perhaps not surprising to see that Equation~(\ref{eqn:TTVp})
and Equation~(\ref{eqn:TTVs}) are intimately connected, via

\begin{align}
\mathrm{TTV}_S(t) &= -\frac{1}{M_{SP}} \mathrm{TTV}_P(t),
\label{eqn:TTVs2}
\end{align}

where the $t$ functionality indicates that this a fully general expression at
all times, and indeed for any choice of inclination angles $i_S$ and
$\Omega_S$. This is very important. If one takes a known TTV signal, which
appears sinusoidal and thus moon-like, one can calculate the location of the
expected moon modulo just one unknown parameter - $M_{SP}$. For a given choice
of $M_{SP}$ then, one can predict where the moon should be and go further and
stack those events to create a moon-folded transit light curve. It should be
emphasised that such moon-folding is only possible if a TTV signal is in hand.

Despite the fact that Equation~(\ref{eqn:TTVs2}) is formally independent of
$i_S$ and $\Omega_S$, recall that an assumption upon which its derivation is
predicated is that the orbital motion of the planet and moon are predominantly
in the $\hat{X}$-direction. As the orbit becomes increasingly inclined, that
assumption will be stressed, making Equation~(\ref{eqn:TTVs2}) increasingly
approximate in nature. And so, in this regime, one would expect moon-folding to
still yield a signal, but attenuated to some degree as a result of this effect,
which is illustrated later in numerical tests (see Section~\ref{sec:tests}).
In such a case, the optimal reconstruction would require a fully photodynamic
algorithm (e.g. \luna; \citealt{luna:2011}). This highlights how moon-folding
is a powerful tool for initial detection but characterisation of the signal
warrants other tools.

\subsection{Bounds on $M_{SP}$}
\label{sub:Msplimits}

For a given exoplanetary TTV signal, one can propose a trial $M_{SP}$ mass
ratio and then perform a light curve fold for a putative exomoon (more details
on the suggested procedure are provided later in Section~\ref{sub:algorithm}).
Given that there is only one unknown, the problem is well-suited for a simple
grid search, similar to how a periodogram used in radial velocity surveys for
example \citep{cumming:2004,gregory:2007,zechmeister:2009}. In that example,
the range of periods to be explored is not unbounded, typically constrained by
the Nyquist frequency and the baseline of observations. Similarly, the
$M_S/M_P$ grid also has a physical boundary between 0 and 1. However,
efficiency can be improved by further tightening these limits as argued in what
follows.

For a given TTV signal, one can measure its amplitude and period
straight-forwardly. The TTV amplitude is proportional to $a_{SB} M_{SP}$
(Equation~\ref{eqn:TTVp}), but the planet-moon separation cannot grow
without limit - orbital stability must be considered. This allows one
to define a limiting value for $M_{SP}$, following a similar logic
to that used recently by \citet{impossible} to define the maximum TTV amplitude
possible due to an exomoon. Let us define the maximum allowed planet-moon
separation, $a_{SP}$, as $a_{SP,\mathrm{max}} = f_{\mathrm{max}}
R_{\mathrm{Hill}}$, where $f_{\mathrm{max}}$ is a value
less than one and $R_{\mathrm{Hill}}$ is the planet's Hill radius.

\begin{align}
a_{SB,\mathrm{max}} =& \frac{a_{SP,\mathrm{max}}}{1+M_{SP}} ,\nonumber\\
\qquad=& \frac{f_{\mathrm{max}} a_P (q/3)^{1/3}}{1+M_{SP}}.
\end{align}

Turning back to the planetary TTV signal in Equation~(\ref{eqn:TTVp}), it is
noted that the inclination terms in parentheses can never exceed unity, and
thus

\begin{align}
\mathrm{TTV}_P \leq& \frac{a_{SB} M_{SP}}{a_P n_P}.
\end{align}

Using this, and the maximum possible planet-moon separation, allows
one to estimate the lighest possible moon which could explain a given TTV
amplitude:

\begin{align}
M_{SP} \geq \Bigg( \frac{ f_{\mathrm{max}} (q/3)^{1/3} }{ n_P \mathrm{TTV}_P } - 1 \Bigg)^{-1}.
\label{eqn:Mspmax}
\end{align}

In principle, $M_{SP}$ could be up approach unity, with the greatest
allowed value occurring when $a_{SP}$ takes the smallest allowed value
(for example the Roche limit of the planet). However, in practice, as
the planet-moon separation shrinks, one of the explicit assumptions in
the TTV theory described in \citet{thesis:2011} becomes invalid. This
occurs when the moon period is comparable to the transit duration,
such that the moon essentially washes out its own TTV. If one requires
$P_S \gg T$, then one can set $P_S \geq g T$, where $g$ is chosen to be
some number greater than unity, with 10 being suggested here (i.e. an
order-of-magnitude). This can be related to planet-moon separation divided
by the Hill radius, $f$, using Equation~(12) of \citet{kipping:2009a}:

\begin{align}
f_{\mathrm{min}} &= \Bigg( 3 \Big(\frac{ g T }{P_P}\Big)^2 \Bigg)^{1/3},
\end{align}

such that

\begin{align}
M_{SP} \leq \Bigg( \frac{ f_{\mathrm{min}} (q/3)^{1/3} }{ n_P \mathrm{TTV}_P } - 1 \Bigg)^{-1}.
\label{eqn:Mspmin}
\end{align}

\subsection{The transit origami recipe}
\label{sub:algorithm}

To illustrate moon-folding, a suggested toy algorithm is now introduced.
However, it is highlighted that reasonable variants of some of the choices made
in what follows would also produce a viable moon folding strategy.

The planetary transit will be deeper than the satellite transit and thus must
be dealt with. The signal could be modelled out, but given how small moons
could be, any slight residual error in that process would potentially
introduce false-positives. Accordingly, it is suggested here that one should
simply exclude the planetary transit altogether. A proposed recipe for
achieving this is now described, which is broken down into three steps (1, 2,
3), which themselves have multiple parts (a, b, c). Beginning with step 1:

\begin{itemize}
\item[{\textbf{1a]}}] Infer the transit time and transit shape parameter
posteriors for all available epochs.
\item[{\textbf{1b]}}] Calculate a 2\,$\sigma$ lower (upper) limit on the time
of planetary first (fourth) contact for each epoch.
\item[{\textbf{1c]}}] For each epoch, take the detrended photometric time
series and exclude all times that occurs within the 2\,$\sigma$ confidence
planetary transit window creating a ``planet-cleansed'' time series.
\end{itemize}

Having excluded the planetary transits, step 2 next finds the points expected
to occur within the moon's transit. In what follows, it is assumed that the
moon transit has approximately the same duration as the planetary transit. This
is suitable when the moon's velocity, is much less than the barycentric
velocity, but will become stressed for moons on tight orbits. For circular
orbits, and again using Equation~(12) of \citet{kipping:2009a}, this assumption
corresponds to $f \gg 3 (q/3)^{2/3}$ - as an example, $q = 10^{-3}$ implies
$f \gg 0.014$. This assumption thus only impacts the most compact moons; moons
which are unsuitable for moon-folding anyway since their short periods
average out their own TTVs (as well producing very small TTVs even if this
wasn't true). Accordingly, let us proceed with the following sub-steps for step
2:

\begin{itemize}
\item[{\textbf{2a]}}] Choose a trial $M_{SP}$ value between the minimum and
maximum allowed values (e.g. as part of a grid search).
\item[{\textbf{2b]}}] Using Equation~(\ref{eqn:TTVs2}), calculate a posterior
distribution for each epoch's mid-transit time for the moon.
\item[{\textbf{2c]}}] Add these posteriors to $\pm0.5$ the transit duration
posteriors (keeping track of mutual covariances), to calculate the expected
time of the satellite's first and fourth contact (for example by taking the
median of the posteriors).
\end{itemize}

A trapezoidal shaped transit's signal-to-noise is maximised by choosing a
duration that corresponds to the FWHM (rather than the first-to-fourth contact
duration; \citealt{carter:2008}) and thus it is suggested to employ that
definition in the above. Finally, in step 3, the planet-cleansed light curve is
separated into two arrays for the in- and out-of-moon transit:

\begin{itemize}
\item[{\textbf{3a]}}] From the cleansed data, select only the times which occur
inside the satellite transits. Create a second array for the other data.
\item[{\textbf{3b]}}] For both arrays, subtract the time of expected moon
mid-transit time at an epoch-by-epoch level.
\item[{\textbf{3c]}}] Combine the epochs together, to create two final arrays;
one for in-moon transit and one for out-of-moon transit.
\end{itemize}

Step 1 need only be done once, but steps 2 \& 3 will be repeated for each
choice of $M_{SP}$. Much like a periodogram, one searches through a grid
of $M_{SP}$ values, creates the corresponding folded light curves, and then
measures statistical quantities pertaining to signal strength and significance.
This can generally be done thousands of times in minutes on a modern
laptop with typical data sets.

For the statistical score quantifying significance/power, it is suggested here
to use a simple box model (similar to \citealt{kovacs:2002}), which is then
compared to the null hypothesis of a flat line. The box's width is fixed to the
planetary transit duration (as used in step 2), and the central time is also
fixed for a given choice of $M_{SP}$, and thus one simply need measure the
weighted average of the photometric points in and out of the putative moon dip.
The null hypothesis' flat line is the weighted sum of all points. A
$\Delta\chi^2$ score can then be used to assess significance. 

\subsection{Possible future recipe improvements}

It is emphasised that the suggested recipe is not the only way to conduct
moon folding, but merely a simple straight-forward one that was found to work
well (see Section~\ref{sec:tests}). For future iterations, some possible
suggested improvements are highlighted.

First, limb darkening is ignored in the above, in the same way as BLS
does\citep{kovacs:2002}. Recent work has found that $\sim10\%$ improvements in
planet search sensitivity can be achieved using a modification of BLS that
accounts for limb darkening, dubbed Transit Least Squares (TLS;
\citealt{hippke:2019}). Similarly, it would be worthwhile to investigate this
here.

Second, for a given guessed $M_S/M_P$ value, a reconstructed signal is obtained
assuming the moon duration is equal to that of the planet. Since that reconstructed
signal ``knows'' the size of the moon (given by the depth), this could be used
to refine the duration slightly - as the smaller moon radius will slightly
reduce the overall duration. This is a small effect but could further improve
sensitivity.

Finally, although the notebooks used to generate the results in this paper are
publicly available (\wwwcoolworlds), a clean, user-friendly open-source
package would be of obvious benefit to the community.

\section{Signal-to-Noise Ratio}
\label{sec:SNR}

The double-folding technique discussed in Section~\ref{sec:theory} provides a
pathway to quickly flagging possible exomoon candidates for exoplanets
exhibiting TTVs. One can extend the formalism to estimate the expected
signal-to-noise ratio (SNR) of an exomoon relative to the planet. In what
follows, an approximate method for estimating this is presented, which should
not be treated as an exact calculation, but rather as an observational guide.

\subsection{Mean duration of the isolated moon transits}
\label{sub:meanduration}

Since the moon folding technique only uses the segments of the exomoon
light curve that occur outside of the planetary transit, one needs to
estimate the mean duration of these segments to make progress. From
Figure~\ref{fig:halfhouse}, one can see that the portion of the exomoon
transit that occurs outside of the planetary transit is given by

\begin{align}
T_{S,\mathrm{out}}(t) &= \mathrm{min}[|\mathrm{TTV}_S(t)| + |\mathrm{TTV}_P(t)|,T]
\label{eqn:TS1}
\end{align}

where the minimum ($\mathrm{min}$) function caps the moon duration at
$T$, the duration to independently cross the stellar disk. As depicted in
Figure~\ref{fig:halfhouse}, it is assumed here (as was done earlier) that the
duration of the exomoon transit is the same as that of the planet, which
implies the orbit is nearly coplanar and that the orbital velocity of the moon
around the barycentre is much less than the orbital velocity of the barycentre
around the star. This is the same assumption used in TTV theory derivations,
specifically it represents assumptions $\alpha3$ and $\alpha4$ in the
derivation presented in \citet{thesis:2011}.

Plugging Equation~(\ref{eqn:TTVs2}) into Equation~(\ref{eqn:TS1}), one finds

\begin{align}
T_{S,\mathrm{out}}(t) &= \mathrm{min}[ |\mathrm{TTV}_P(t)| (1+M_{SP}^{-1}),T].
\label{eqn:TS2}
\end{align}

To calculate the mean duration over all times, one needs to integrate the
above. The TTV waveform becomes sinusoidal is the limit of $\Omega_S \to 0$
(see Equation~\ref{eqn:TTVp}) which is adopted as an assumption in what follows
to simplify the calculation. Replacing time, $t$, with orbital
phase, $\phi \equiv n_s (t-\tau_s)$, one finds $\mathrm{TTV}_P =
-A_{\mathrm{TTV-P}} \sin(\phi)$, and thus

\begin{align}
<T_{S,\mathrm{out}}> &= \frac{\int_{0}^{2\pi} T_{S,\mathrm{out}}(\phi) \mathrm{d}\phi}{\int_{0}^{2\pi} \mathrm{d}\phi}.
\end{align}

The mean duration should be compared to the duration of the planetary event
for context, and so it is useful to define the duration ratio
$\mathcal{D} \equiv \frac{<T_{S,\mathrm{out}}>}{T}$, which can be shown to
satisfy

\begin{equation}
\mathcal{D}(\Lambda) = \frac{2}{\pi}
\begin{cases}
\Lambda & \text{if } 0 < \Lambda \leq 1 ,\\
\Lambda - \sqrt{\Lambda^2-1} + \mathrm{sec}^-1[\Lambda] & \text{if } \Lambda > 1 ,
\end{cases}
\end{equation}

where $\Lambda$ is defined as

\begin{align}
\Lambda \equiv A_{\mathrm{TTV-P}} (1+M_{SP}^{-1})/T,
\label{eqn:Lambda}
\end{align}

and $A_{\mathrm{TTV-P}} \equiv (a_{SB} M_{SP})/(a_P n_P)$.

\subsection{Effective SNR of the moon folded dip}
\label{sub:SNR}

Section~\ref{sub:meanduration} establishes and formulates that the
moon's isolated transit has a mean duration equal to some fraction
of the full planetary duration. The SNR of this signal, relative to
the planetary signal, will be scaled by the square root of this
duration ratio, as well as the transit depth ratio. This can be
solved for analytically under the simplifying assumptions of the
expressions presented in this section.

Before doing so, it useful to re-express $\Lambda$ in terms of
the exomoon's physical parameters, rather than the planetary TTV
amplitude. Expanding the equation for $A_{\mathrm{TTV-P}}$, one has

\begin{align}
A_{\mathrm{TTV-P}} &= \frac{ a_{SB} M_{SP} }{ a_P n_P },\nonumber\\
\qquad&= \frac{ a_{SP} }{ a_P n_P } \frac{1}{M_{SP}^{-1}+1}
\end{align}

And plugging this into Equation~(\ref{eqn:Lambda}), one finds the
$M_{SP}$ terms cancel to give

\begin{align}
\Lambda &= \Big(\frac{1}{2\pi}\Big) \Big(\frac{ a_{SP} }{ a_P }\Big) \Big(\frac{P}{T}\Big).
\label{eqn:Lambda}
\end{align}

As an alternative formulation, the $a_{SP}$ term can be replaced in the
above with a moon located at $x$ planetary radii.

\begin{align}
\Lambda &\simeq \Big(\frac{x p}{2\sqrt{1-b^2}}\Big),
\end{align}


The SNR of the exomoon folded signal is now

\begin{align}
\mathrm{SNR}_S &= R_{SP}^2 \sqrt{\mathcal{D}(\Lambda)} \mathrm{SNR}_P.
\label{eqn:SNRs}
\end{align}


\subsection{Comparison to OSE}

We note that the SNR of exomoon transits for the portion of the light curve
outside of the planetary transit will be approximately the same whether one
uses the moon folding method (this work) or OSE \citep{heller:2014}. There's no
``free lunch'' in an SNR-sense by simply shuffling the transits around in
different ways (although small differences will occur due to approximations in
the transit shape). Photodynamical modelling (e.g. \luna; \citep{luna:2011})
will, in general, always lead to higher SNRs since it includes numerous effects
ignored with both approaches; such as variable moon transit durations as
a function of orbital phase, light curve asymmetry due to acceleration effects,
etc.

Although OSE and moon folding produce similar SNRs, a clear advantage of moon
folding is that a coherent signal is reconstructed. This allows individual
epochs to be inspected for consistency with the stacked signal. One
can also freely median-bin the folded light curve, producing a composite
signal robust to outlier measurements - which is not possible for OSE
due to the duty cycle effect discussed earlier. Moon folding also features
a simple search statistic which can be sought in a 1D grid, akin to a
BLS periodogram. Finally, the moon folding method recovers not only
the moon's relative size, but also its relative mass, allowing for an extra
layer of statistical testing by comparison to physical models. On the
downside, moon folding only works if a TTV signal is present, unlike OSE.
However, high confidence exomoon detections will ultimately demand both
dynamical and transit signatures regardless. Yet more, it may be possible
to extend this method to non-TTV systems as discussed in
Section~\ref{sec:discussion}.

\subsection{SNR calculation for the \kepler\ catalog}

To provide a sense of scale, Equation~\ref{eqn:SNRs} was evaluated on
the \kepler\ catalog. Rather than curate TTVs for each planet and attempt
any kind of model selection for their potential to be a moon, let us instead
simply use the $\Lambda$ formula of Equation~(\ref{eqn:Lambda}). For the
satellite, the Solar System moons provide a plausible set of input parameters
for this calculation. Doing so, it was found that Ganymede consistently led
to the most favourable SNRs and thus is adopted exclusively in what follows.

For some short-period planets, the smaller Hill spheres would place Ganymede
(located at 15 planetary radii) outside of the stable orbital range. In such
cases, the distance was truncated down to half a Hill sphere, where planetary
masses are estimated using \forecaster's conservative lower bound
\citep{chen:2017}.

Only five \kepler\ planetary systems have predicted SNRs greater than 3 for a
Ganymede moon:
Kepler-93b ($\mathrm{SNR}_S = 3.5$),
KOI-314b ($\mathrm{SNR}_S = 3.3$)
Kepler-142b ($\mathrm{SNR}_S = 6.1$),
Kepler-142c ($\mathrm{SNR}_S = 5.8$)
and Kepler-142d ($\mathrm{SNR}_S = 5.5$).
KOI-314b experience a near mean-motion resonance with KOI-314c
\citep{kipping:2014} and thus although it exhibits TTVs, this would greatly
complicate any attempt to extract a moon-only TTV component. Kepler-142 is an
extremely compact system around a late M-dwarf \citep{muirhead:2012}, in many
ways a precursor discovery to TRAPPIST-1 \citep{gillon:2017}. As a result,
exomoons are improbable due to dynamical packing \citep{kane:2017}. This leaves
the small, rocky planet Kepler-93b \citep{ballard:2014}, which could be worth
future investigation although no TTVs are presently known for the system
\citep{dulz:2017}.

\section{Numerical Tests}
\label{sec:tests}

\subsection{An example system}
\label{sub:example}

\begin{figure*}
\begin{center}
\includegraphics[width=18.0cm,angle=0,clip=true]{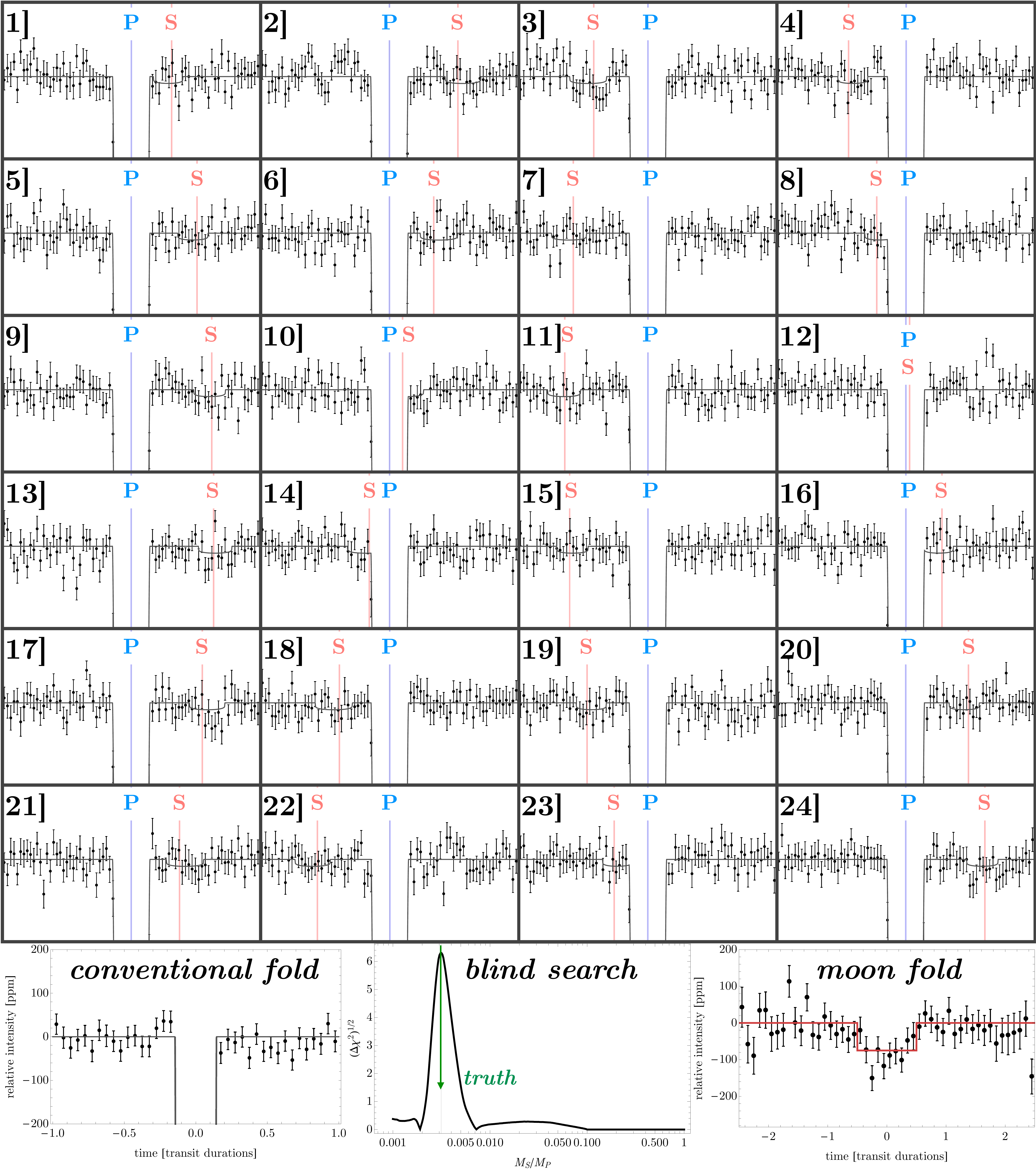}
\caption{
Panels 1] to 24] show a zoom-in of the 24 transit epochs simulated of the
hypothetical Jupiter+Earth around a Sun-like star described in
Section~\ref{sub:example}. The blue vertical lines mark the
location of the planetary mid-transit time (``P'') , whereas the red mark the
predicted location of the satellite (``S'') transit using
Equation~(\ref{eqn:TTVs2}) and using the blindly recovered mass ratio,
$M_S/M_P$. Bottom-left shows a simple linear fold of the transits,
which is incoherent for the moon events thus making them indistinguishable
from the noise. Bottom-middle shows a grid search of the $\Delta\chi^2$
significance of moon fold trials as a function of $M_S/M_P$, with a clear
peak at the true value. Bottom-right shows the moon fold corresponding to
the most significant $M_S/M_P$ value.
}
\label{fig:example}
\end{center}
\end{figure*}

Before discussing a grid of tests that were performed, it is useful to begin
with a single example for illustrative purposes. Consider a hypothetical test
system comprising of a Jupiter-sized/mass planet orbiting a Sun-sized/radius
star on a circular path with an orbital period of 60\,days. The planet transits
the star with an impact parameter of $0.5$, and the star has a quadratic limb
darkening profile of $q_1=q_2=0.3$ \citep{q1q2}. Around this planet orbits a
single Earth-sized/mass moon at the Hill radius ($f=1$) and in a coplanar,
circular orbit. This means that the planet is ${\sim}317$ times more massive
than the moon, such that $M_S/M_P = 0.00315$.

A baseline of 4\,years of photometric data was assumed, similar to that of the
\kepler\ Mission. The light curves are generated using the photodynamic
\luna\ algorithm for planet-moon transits \citep{luna:2011}. Uncorrelated
Gaussian noise is then added to the simulated light curves; specifically, the
noise is equivalent to an RMS of 52.7\,ppm over 6 hours (similar to the median
6-hour CDPP for \kepler\ dwarfs 54.5\,ppm; \citealt{christiansen:2012}), or
1\,mmag over a minute. In total, 24 transits were generated and are shown in
Figure~\ref{fig:example}.

A grid of 301 $M_S/M_P$ values were generated, separated uniformly in log-space
from $10^{-3}$ to $1$. This range extends beyond the bounds described in
Section~\ref{sub:Msplimits}, in order to provide a full inspection of the
parameter space. We follow the recipe in Section~\ref{sub:algorithm} and
measure the $\Delta\chi^2$ between a simple box-model and a flat-line at each
$M_S/M_P$ trial. Since the simulated data are uniformly sampled, it is possible
to derive the planetary transit time for each epoch non-parametrically by using
a flux weighted centroid.

A strong, unimodal peak emerges at $M_{SP} = 10^{-2.5} =
0.00316$ (see Figure~\ref{fig:example}), almost exactly the same value as the
truth. For the assumed noise, this is a $\Delta\chi^2=39.7$, or
$\sqrt{\Delta\chi^2}=6.3$\,$\sigma$ detection. It is noted that using the true
photodynamic model only marginally improves this to $7.7$\,$\sigma$. In
other words, the assumptions inherent to moon-folding method means that it
recovers 82\% of the maximum possible SNR. This demonstrates how the moon-folding
technique is generally expected to recover the large majority of the theoretical
maximum SNR.

\subsection{Varying the moon parameters}
\label{sub:Nexamples}

Let us next consider repeating the above but varying three of the exomoon
parameters, in order to investigate their impact on the sensitivity of
the moon-folding technique. This also provides an opportunity to investigate
the overall changes in SNR as moon parameters are varied, since different
parameter inputs will affect the amount of time the moon spends in transit
\citep{martin:2019}. To this end, the first two parameters varied describe
the orbital orientation, $i_S$ and $\Omega_S$, which are varied in $30^{\circ}$
intervals. A third parameter that is varied is $f$ in 0.1 steps, which
corresponds to the exomoon semi-major axis in Hill radii. Since semi-major axis
and orbital period are related via Kepler's Third Law, then this also changes
the orbital period of the moon.

\begin{figure*}
\begin{center}
\includegraphics[width=18.0cm,angle=0,clip=true]{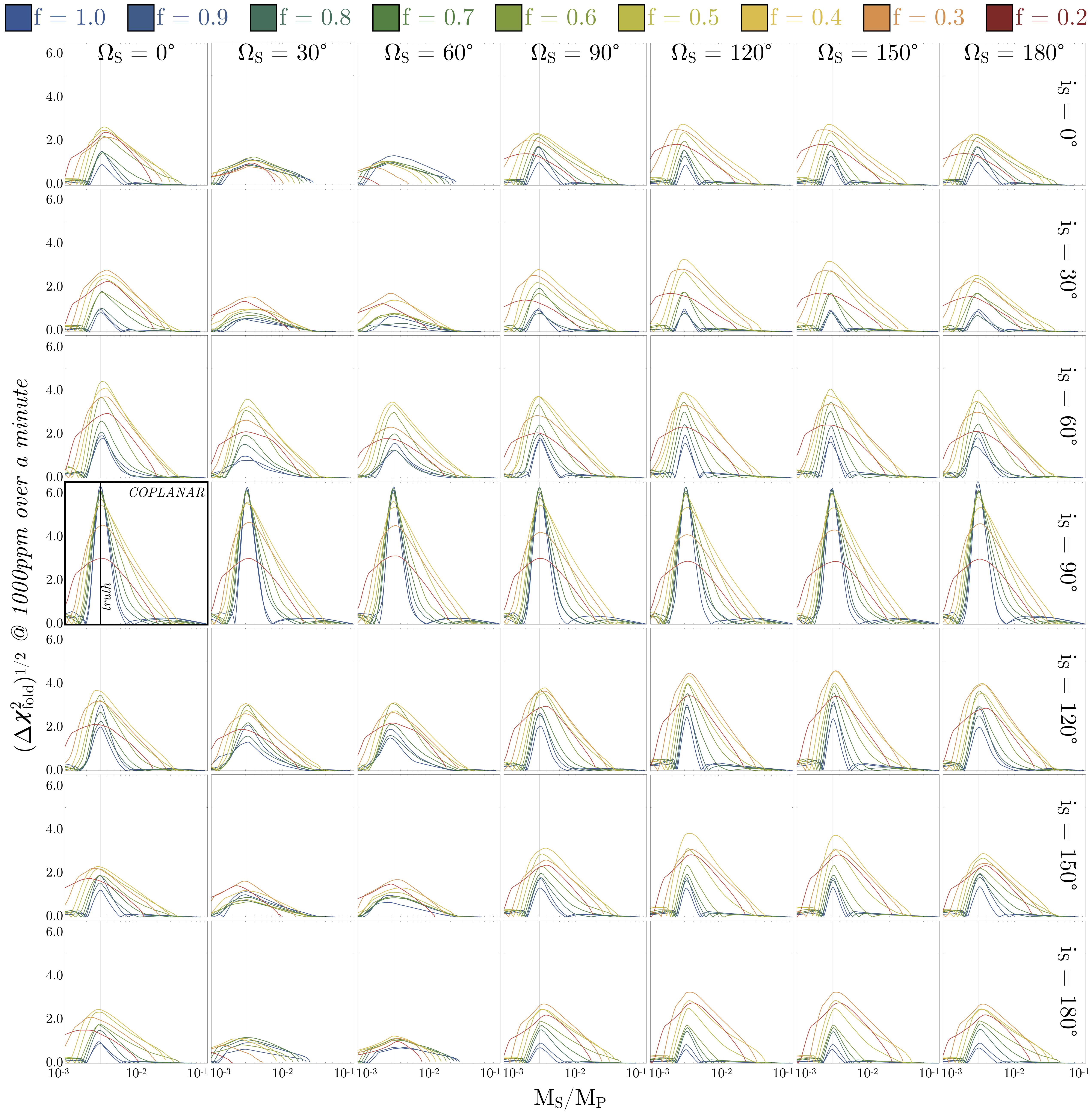}
\caption{
A multi-panel figure showing the effect of various choices of $i_S$ (rows),
$\Omega_S$ (columns) and $f$ (colours; see top) on the moon-folding grid
search. Each panel depicts the absolute $\sqrt{\Delta\chi^2}$ between the moon
box model and a flat-line, which consistently recovers a signal at the
correct $M_S/M_P$ value of $0.00315$, with the exception of some close-in
moons in extremely inclined orbits. The detectability of the moon decreases
in inclined configurations since it simply transits less often (see
\citealt{martin:2019}).
}
\label{fig:abs}
\end{center}
\end{figure*}

\begin{figure*}
\begin{center}
\includegraphics[width=18.0cm,angle=0,clip=true]{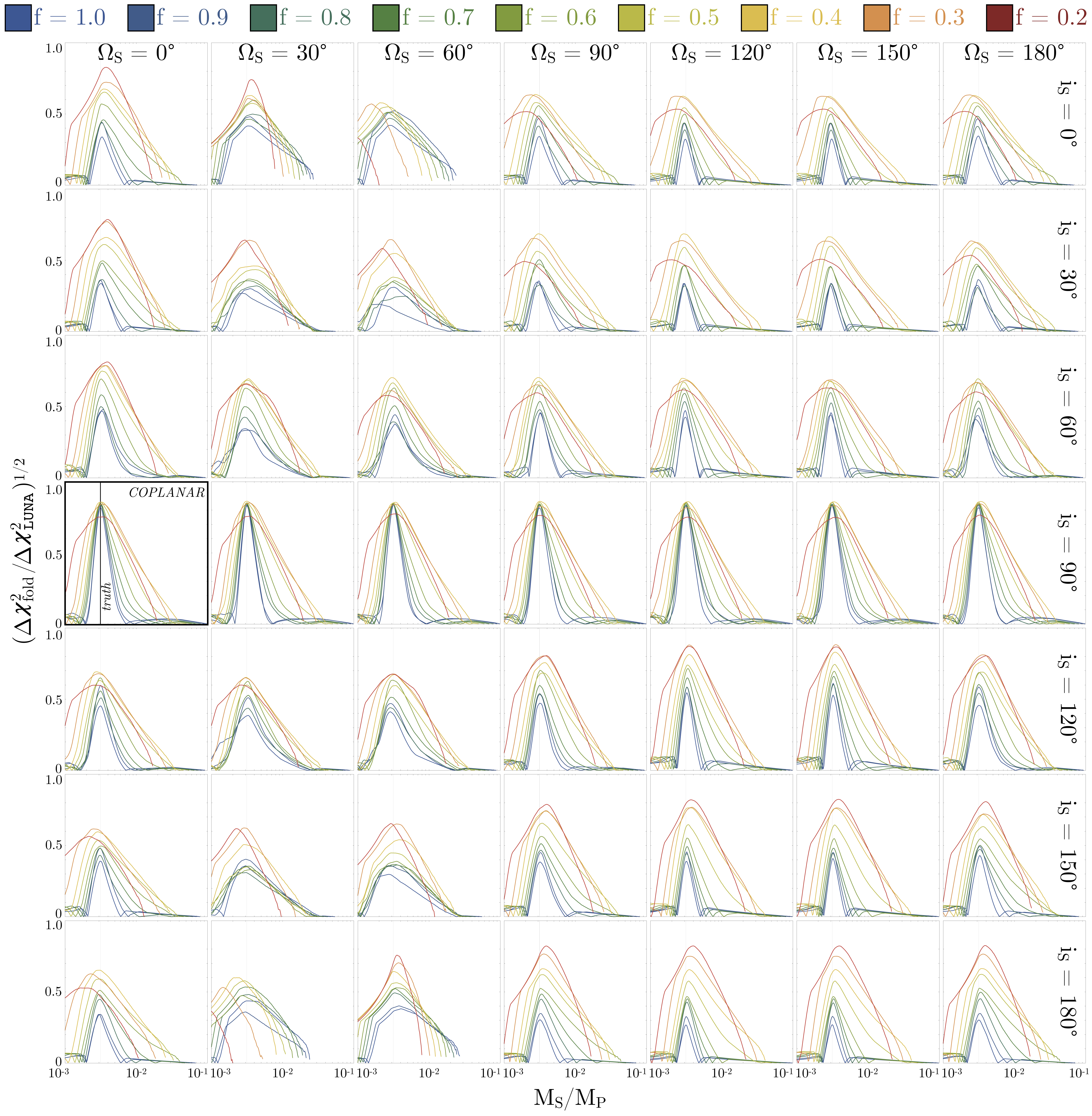}
\caption{
Same as Figure~\ref{fig:abs} except the change in $\chi^2$ is normalised
to that produced when using the true photodynamic model used to generate
the systems (\luna). The moon-folding technique generally recovers over
half of the available signal and for coplanar configurations it exceeds
80\%.
}
\label{fig:rel}
\end{center}
\end{figure*}

In each case, the procedure described in the previous subsection is repeated
and used to produce two plots (Figures~\ref{fig:abs} \& \ref{fig:rel}). The
first is the absolute $\Delta\chi^2$ change with the same assumed noise level
as before, which provides an estimate of detectability and is shown in
Figure~\ref{fig:abs}. The second is the relative change in $\chi^2$ versus that
found from the true photodynamic model (but using only the same planet-cleansed
photometric points). This relative quantity is a useful gauge of how much of
the full signal the moon-folding is able to recover and is independent of the
assumed noise level (see Figure~\ref{fig:rel}). Not surprisingly, the closest
moon orbit considered in the $f$-grid, $f=0.1$, comes into tension with the
assumption of a slow-moving moon and thus is not used.

\subsection{Interpreting the simulated results}
\label{sub:interpretation}

The first key result from this exercise is that in almost all cases the peak
of the $M_S/M_P$ blind search centres on the expected true value, with the only
exceptions being compact moons in nearly perpendicular geometries. On this
basis, one should expect that the moon-folding technique will generally produce
an accurate estimate of the most probable $M_S/M_P$ value for white noise
dominated light curves.

The second key result is that the detectability of exomoons appears to
diminish for inclined configurations, as shown in Figure~\ref{fig:abs}. This is
not surprising and simply a result of the lower duty cycles of such systems
\citep{martin:2019}. This is also supported by the fact that
Figure~\ref{fig:rel} shows far less attenuation of the inclined states, because
even the photodynamic true signal is less detectable. This stems from the fact
that inclined orbits will not necessarily transit each time and thus this will
diminish the detectability of the stacked signal. In other words, in some of
the stacked epochs, there is no dip to fold in. For coplanar moons, then, the
moon-folding strategy produces a fully coherent signal, but for inclined states
the signal can become ``semi-coherent''.

The third key result is that, from Figure~\ref{fig:rel}, the moon-folding
technique recovers typically at least half of the available SNR
but up to 80\% for coplanar configurations. For a large moon which forms from
the circumplanetary disk, such as Titan around Saturn, coplanar configurations
are the natural outcome \citep{canup:2002,inderbitzi:2020} and thus these
should be well-recovered with this strategy.

Finally, it is noted that for coplanar configurations, wider moons are
generally more detectable (by overlapping less often with the planet) and the
moon-folding technique's recovered signal, when compared to \luna, is broadly
independent of $f$. However, for inclined states, wide moons become less
detectable because they have an increasingly large chance to not transit the
star each epoch and thus evade detection \citep{martin:2019}. In a relative
sense, the moon-folding approximation also becomes worse for such inclined
moons because their paths deviate away the nominal $\hat{X}$-chord ever more.

\section{Application to an Example System}
\label{sec:application}

\subsection{Selection of Kepler-973b}

As a final way of testing the new algorithm, a suitable \kepler\ transiting
planet was sought as a real example case. A basic requirement of the new method
is a planet which exhibits TTVs. Initially, Kepler-1625b might seem to be an
ideal test case, given the known exomoon candidate there \citep{teachey:2018}.
However, with four transits, a linear ephemeris + sinusoid model is
under-constrained \citep{undersampling} and thus it is not possible to use the
moon-folding method since one can't make unique predictions for the satellite's
position. Looking further afield, TTVs are common within the \kepler\ catalog
and so some thought on how to choose an appropriate system is required.

Following the aliasing theory presented in \citet{undersampling}, 90\% of
exomoons are expected to produce TTVs of periods between 2 to 20 cycles, and
thus a known TTV system in this range was sought. Further, excessively large
TTVs, or moderately large TTVs lacking associated TDVs, can be flagged as
so-called ``impossible moons'' following the approach of \citet{impossible}.

With these filters in mind, let us take the \citet{ofir:2018} ``spectral'' TTV
catalog as a starting point, which tabulates \kepler\ exoplanets exhibiting
periodic TTVs to a false-alarm probability of $<1$\%. From these, TTV periods
$>20$\,cycles were first filtered out \citep{undersampling}. Next, the
surviving sample was cross-matched against those exoplanets identified in
\citet{impossible} that exhibit a periodic TTV at a statistical threshold
of $\Delta(\mathrm{BIC})>10$ (using the \citet{holczer:2016} TTV catalog)
and excludes any impossible moon cases.

Finally, the 24 cases were ranked by predicted $\mathrm{SNR}_S$ using
Equation~(\ref{eqn:SNRs}) and assuming a Ganymede-analog moon. This
presented two \kepler\ planets with $\mathrm{SNR}_S(\mathrm{Ganymede}) \sim 1$,
noteably Kepler-858b ($\mathrm{SNR}_S(\mathrm{Ganymede}) = 1.1$) and
Kepler-973b ($\mathrm{SNR}_S(\mathrm{Ganymede}) = 0.8$). Of the two,
Kepler-973b exhibits a sinusoidal, moon-like TTV pattern, whereas Kepler-858b
was more saw-tooth like. Accordingly, in this example, let us proceed with
Kepler-973b, a $49.6$\,d $2.1$\,$R_{\oplus}$ single-planet system around
a $K_P = 13.2$ K3-dwarf.

\subsection{Data processing}
\label{sub:dataproc}

The long-cadence (no short-cadence was available) time series \kepler\
photometry was detrended using the method marginalised approach described in
\citet{teachey:2018}. The 28 transit epochs were split into three 10-9-9
groups which were fitted with a \citet{mandel:2002} transit model allowing
each epoch to have its own free transit time, $\tau_i$. This splitting reduces
the total number of free parameters in each regression to a manageable size
yet allows the transit shape constraint from other epochs to improve the
obtainable precision in $\tau_i$ (see \citealt{teachey:2017} for the first
example of this in action). A freely fitted quadratic $q_1$-$q_2$ limb
darkening law was used \citep{q1q2} throughout, with the quarter-to-quarter
contamination values accounted for \citep{seager:2003},
and numerical resampling was implemented on the long-cadences to account
for integration time following the method of \citet{binning:2010}. Regressions
were performed using \multi\ \citep{feroz:2009}.

\begin{figure}
\begin{center}
\includegraphics[width=8.4cm,angle=0,clip=true]{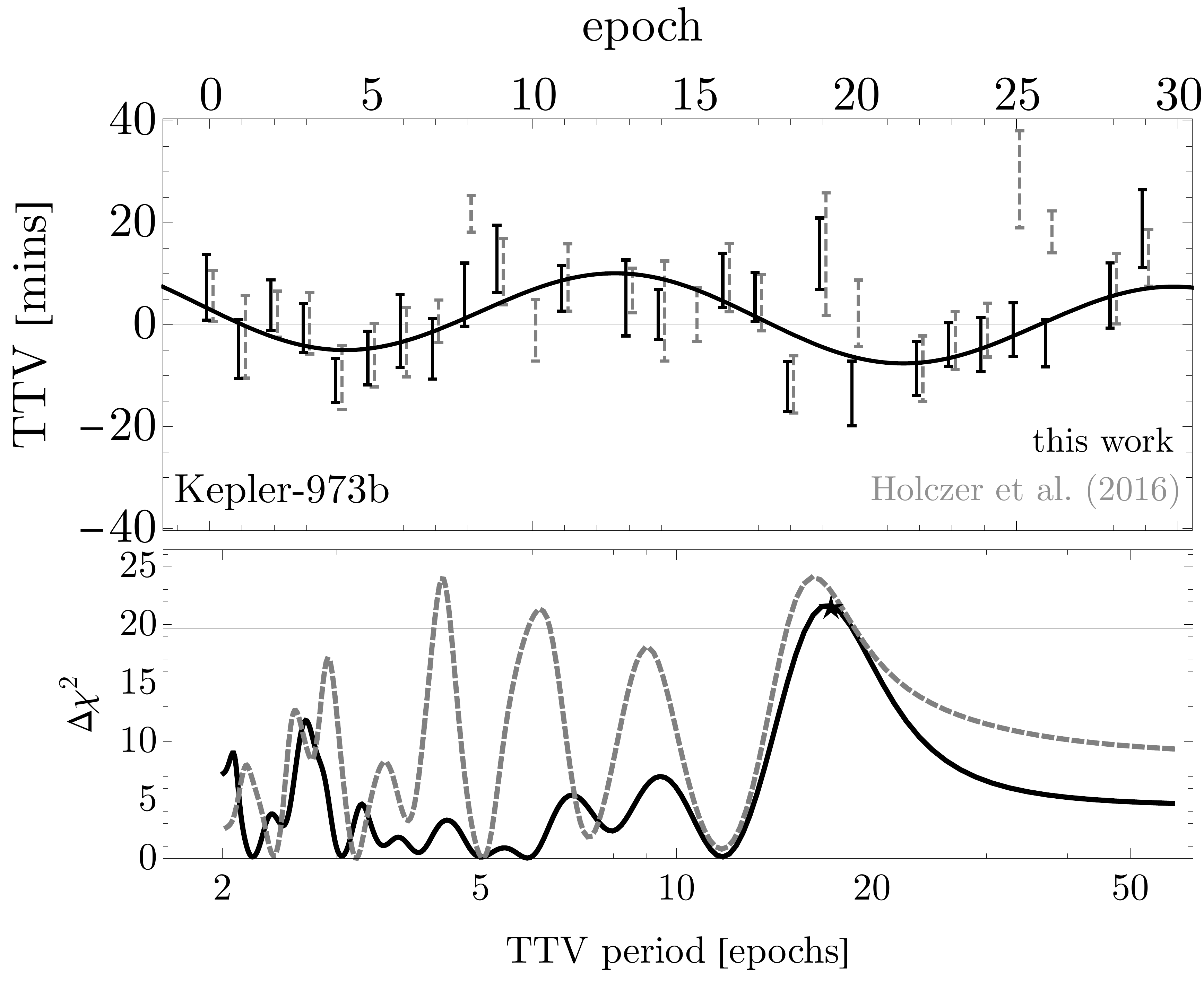}
\caption{
\textbf{Top:} TTVs of Kepler-973b derived in this work (black)
versus those of \citet{holczer:2016} (gray, dashed). Overlaid is
the best-fitting sinusoid to the TTV data of this work.
\textbf{Bottom:} Lomb-Scargle periodograms of the above TTVs,
where the horizontal line marks $\Delta(\mathrm{BIC})=10$.
}
\label{fig:K973b_TTVs}
\end{center}
\end{figure}

The resulting transit times are shown in Figure~\ref{fig:K973b_TTVs}, where
the derived \textit{a-posteriori} marginalised 1-$\sigma$ credible
intervals (black) are compared with those reported in \citet{holczer:2016}
(grey, dashed). In both cases, the maximum \textit{a-posteriori} linear
ephemeris (found from a separate global fit with no TTVs) has been subtracted.
In addition, the panel below shows the resulting Lomb-Scargle periodogram.
\citet{ofir:2018} report a TTV period of $17.1$\,cycles, which is close to
the best-fit solution in this work of $17.3$\,cycles (amplitude 8.2\,mins).
The results presented here also agree with the \citet{holczer:2016} TTVs, but
generally appear less susceptible to outlier measurements (likely as a result
of the Bayesian segmenting procedure used; \citealt{teachey:2017}).

\subsection{Moon folding}

Let us now proceed following the algorithm detailed in
Section~\ref{sub:algorithm}. In step 2b], one requires an input for
$\mathrm{TTV}_P(t)$. In principle, this could be simply the TTV measurements
directly, but given the sinusoidal nature of the TTVs (which is expected for an
exomoon; see Equation~\ref{eqn:TTVp}), this works uses the best-fitting
sinusoid determined in Section~\ref{sub:dataproc}, (the black line plotted in
the upper panel of Figure~\ref{fig:K973b_TTVs}). To ensure no part of the real
planetary transits leak into the predicted moon windows (which could cause a
false-positive), both the measured and predicted locations of the planetary
transits were excluded.

\begin{figure*}
\begin{center}
\includegraphics[width=18.0cm,angle=0,clip=true]{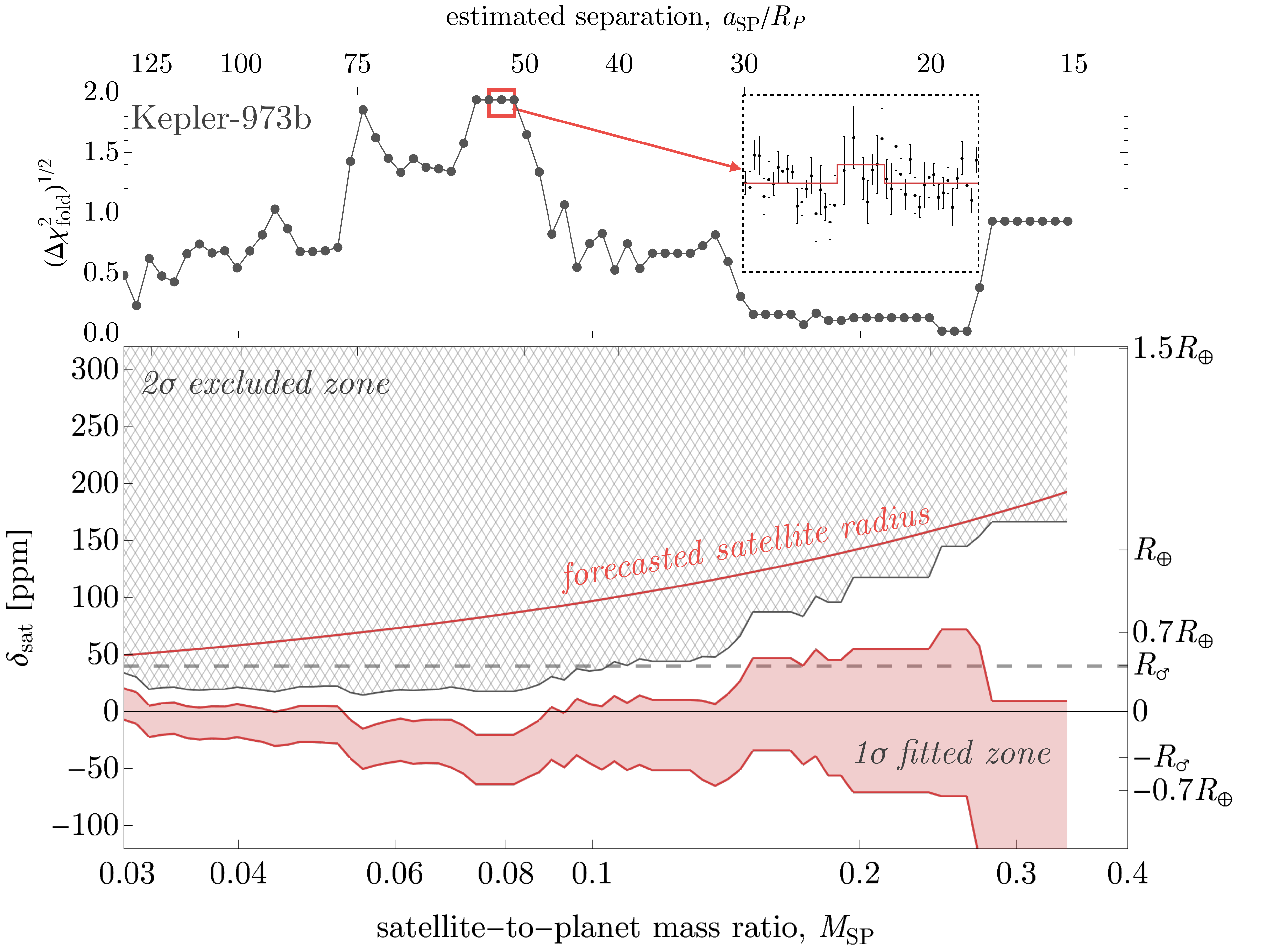}
\caption{
A real example of our moon-folding algorithm on Kepler-973b. Top panel
shows the $\chi^2$ improvement of including a moon-dip versus $M_{SP}$,
revealing no significant dips, with the largest dip being inverted
(inset). Lower panel shows the corresponding transit depths (left-axis)
and moon radii (right-axis), highlighting both the 1\,$\sigma$ fitted
region with solid shading and the excluded zone (hatched). Although no
moon dips are found, our limits are sensitive down to Ganymede-sized
moons.
}
\label{fig:K973b_fold}
\end{center}
\end{figure*}

Using \forecaster\ \citep{chen:2017} and the $R_P/R_{\star}$ posteriors, a
posterior distribution for the planet-to-star mass-ratio, $q$, was inferred.
This was then used with Equations~(\ref{eqn:Mspmax}) \& (\ref{eqn:Mspmin}) to
define a $M_S/M_P$ grid from $0.0297$ to $0.748$. The next step was to scan
along this grid at 100 log-uniformly separated intervals, which took 1\,minute
on a typical desktop computer\footnote{A denser 1000 grid was also tried, but
revealed no other features not apparent in Figure~\ref{fig:K973b_fold}.}. The
results are summarised in Figure~\ref{fig:K973b_fold}, where no significant
moon-like dips were found. The highest significance solution occurs at around
$M_{SP} = 0.08$ yielding a $\sqrt{\Delta\chi^2}=2$ improvement - given the
extra complexity of the moon model, such a modest improvement is not
significant. Indeed, it is also noted that this region corresponds to an
inverted moon transit of $(-42\pm22)$\,ppm and thus can be discounted as a
possible moon (depicted in the inset of Figure~\ref{fig:K973b_fold}'s upper
panel).

Upper limits on the maximum positive moon dip were derived by simple
$\chi^2$ perturbation, seeking a dip which increases the $\chi^2$ versus
the null model of $+4$ (corresponding to 2\,$\sigma$). This is shown by
the hatched region of Figure~\ref{fig:K973b_fold}'s lower panel. The 
right-hand axis converts these transit depths into physical radii by
using the stellar radius, revealing that our observational limits probe
down to 0.4\,$R_{\oplus}$ - which is smaller than Ganymede.

Recall that a small $M_{SP}$ solution corresponds to a wider separated moon,
since it is the product of these two that governs the measured TTV amplitude.
In fact, one can estimate the corresponding exomoon semi-major axis in units
of the planetary radius by re-arranging

\begin{align}
&A_{\mathrm{TTV-P}} = \frac{a_{SB} M_{SP}}{a_P n_P},\nonumber\\
&\Bigg(\frac{a_{SP}}{R_P}\Bigg) = A_{\mathrm{TTV-P}} \Bigg(\frac{M_{SP}}{1+M_{SP}}\Bigg)
\Bigg(\frac{ 2\pi (a/R_{\star}) }{ P_P (R_P/R_{\star}) }\Bigg),
\end{align}

which is used to populate the ticks on the upper $x$-axis of
Figure~\ref{fig:K973b_fold}, displaying a range from 15 to 125 planetary radii.
Below 15 planetary radii, or equivalently above $M_SP > 0.35$, the moon
transits become so close to the planet that the number of points inside the
folded moon signal is less than one datum. For this reason, even though a
higher mass ratio is physically possible via Equation~(\ref{eqn:Mspmax}), our
moon-folding approach is not sensitive to it.

Finally, it is noted that the $M_{SP}$ values on $x$-axis can be used to
predict the expected physical size of such a moon. This was achieved by first
forecasting the planetary mass (since no measurement currently exists) with
\forecaster, then multiplying this by the given $M_{SP}$ value to get a moon
mass, and then forecasting back into radius space. This forecast is shown with
the curved red line in the lower panel of Figure~\ref{fig:K973b_fold}.
Critically, this forecast is inside the observationally excluded zone across
the entire range, meaning that the moon hypothesis can be fully excluded. In
summary, one can use this technique to show that Kepler-973b's TTVs are not
being caused by a single moon.

\section{Discussion}
\label{sec:discussion}

In this work, a new method has been proposed for performing a double fold of
photometric time series of transiting planets in order to search for exomoons.
In what follows, some of the limitations of this approach are discussed, as
well possibles areas for improvement.

The method assumes a slow-moving moon, such that the moon's transit
duration is approximately equal to that of the planet. This is also
an assumption built into the theory of TTVs due to
exomoons\citep{thesis:2011}. Slow-moons also avoid the possibility 
of the moon conducting multiple half-orbits/transits during
the planetary transit, which is not modelled. Further, fast-moons (those
on compact orbits) are so close to the planet that the portion of the
light curve where the moon is transiting but the planet is not is greatly
diminished, minimising the available signal. We estimate a rigorous
condition of $f\equiv(a_S/R_H) \gg 3^{1/3} (M_P/M_{\star})^{2/3}$, and in
practice expect the method to fail at $f \lesssim 0.1$. We also assume
a coplanar moon orbit, but find the method still works for even highly
inclined configurations but at reduced sensitivity - since the moon
spends less time transiting the star \citep{martin:2019}.

An important limitation of the new method is that it only works if the
exoplanet exhibits TTVs. The planetary TTV essentially resolves the moon's
aliased period and phase, leaving just the separation (although in practice
the mass-ratio is used) as the remaining free parameter. In principle, one could
imagine extending this method to non-TTV systems, but the grid search would
necessarily become three-dimensional now, with the moon's phase, period and
separation all being unknown quantities. The latter two parameters could be
combined into a single-term in the case where the planetary mass is precisely
known, although this is somewhat complicated by the fact that methods such as
radial velocities in fact measure the planet+satellite system combined mass.
Nevertheless, this highlights the possibility of extending the transit origami
approach to a broader population in future work.

In general, moon-folding will always be less sensitive than a full
photodynamical model \citep{luna:2011}. By culling the data inside the transit,
approximating that the moon's impact parameter is the same as that of the
planet, and ignoring the acceleration effects of the moon's orbit, the
algorithm presented here cuts several corners in the name of efficiency. The
true value of the approach here is to flag interesting signals, and it will
still be necessary to perform more detailed tests and ensure physical orbital
solutions exist (e.g. see \citealt{teachey:2018}).

An important step in the outlined algorithm is to first mask the known
planetary transits and here a conservative approach is recommended, taking both
the measured transits times and modelled TTV waveform as guides to ensure the
transit is fully removed (as was done in the example application in
Section~\ref{sec:application}). This is crucial to ensure that the wings of the
planetary transit do not induce spurious moon-like signatures. As a result of
this process though, close-in moons may get truncated. Yet more, standard
exomoon TTV theory (and thus this method) breaks down for moons with periods
comparable to the transit duration, which imposes an additional (but unrelated)
barrier to detecting close-in moons. Together then, close-in moons will remain
challenging with the outlined method. For example, in the case of Kepler-973b,
it is not possible to probe interior to 15 planetary radii, which is the
approximate position of Ganymede around Jupiter (i.e. a Callisto would be fine,
but a Europa or Io is too close-in).

To conclude, it is highlighted that exomoons remains a vibrant intellectual
topic. When it comes to detection, new approaches and methodological
developments continue to arise by many teams, and enormous opportunities await
the community when these are coupled to the rise of ever larger exoplanet
surveys.

\section*{Acknowledgments}

The author thanks the anonymous reviewer for their very constructive comments.

The author acknowledges support from NASA Exoplanet Research Program grant number 80NSSC21K0960.

The Cool Worlds Lab is supported by our generous donors, including Tom Widdowson, Mark Sloan, Laura Sanborn, Douglas Daughaday, Andrew Jones, Elena West, Tristan Zajonc, Chuck Wolfred, Lasse Skov, Alex de Vaal, Jason Patrick-Saunders, Methven Forbes, Stephen Lee, Zachary Danielson, Vasilen Alexandrov, Chad Souter, Marcus Gillette, Tina Jeffcoat, Jason Rockett, Scott Hannum, Tom Donkin \& Mark Elliott.

This paper includes data collected by the \textit{Kepler Mission}. Funding for the \textit{Kepler Mission} is provided by the NASA Science Mission directorate.

\section*{Data Availability}

Data used in this paper is from the \textit{Kepler Space Telescope} and can be found at the Mikulski Archive for Space Telescopes (MAST) which is hosted by the Space Telescope Science Institute (STScI). Additional data underlying the analysis was taken from the \citet{holczer:2016} catalogue, which is available on line\footnote{\href{ftp://wise-ftp.tau.ac.il/pub/tauttv/TTV/ver_112}{ftp://wise-ftp.tau.ac.il/pub/tauttv/TTV/ver\_112}}. Notebooks used to generate the figures are made available at \wwwcoolworlds.

\textit{Software used:} \forecaster\ \citep{chen:2017}, \multi\ \citep{feroz:2009},
\luna \citep{luna:2011}, Mandel-Agol light curve model \citep{mandel:2002}.


%

\bsp
\label{lastpage}
\end{document}